# Theoretical demonstration of mode transmission in ZGP-based micrometer waveguide platforms

**Siyi Lu[1], Bo Hu[1,2], Xuemei Yang[1], Yang Li[1], Han Wu[1] and Houkun Liang[1,3]**

*1 College of Electronics and Information Engineering, Sichuan University, Chengdu, Sichuan 610064, China*
*2 hubo_uestc@hotmail.com*
*3 hkliang@scu.edu.cn*

**Abstract:** Birefringence phase-matching based $\chi^{(2)}$ ZnGeP$_2$ (ZGP) waveguide platform has been recently reported for excellent mid-infrared laser generation. Here, a detailed theoretical characterization of mode transmission taking waveguide anisotropy and substrate material absorption into account in a micrometer ZGP waveguide platform (ZGP-on-SiO$_2$) is conducted. Benefited from high-index contrast between ZGP and substrate (SiO$_2$/Air), Transverse electric and magnetic (TM and TE) mode transmission loss at interested wavelengths range of 2 - 12 μm is calculated to be less than 4 dB/cm and 1.5 dB/cm, respectively, in the designed ZGP waveguide. Notably, non-obvious oscillation of mode transmission loss versus phase-matching angles is observed, which is different from that in the previously reported weakly guided anisotropic waveguide. A vital phenomenon named mode crossing at some wavelengths in TM polarization is also exhibited in our waveguide platforms, which jeopardizes waveguide performances and could be avoided by changing the phase-matching angle in practice. This work provides a significant indication of ZGP waveguide design optimization in future and also exhibits extendibility to other birefringent crystal waveguide platforms.

## 1. Introduction

Mid-infrared (MIR, 2.5-25 μm) spectroscopy is widely studied for encompassing multiple atmospheric windows, covering the primary absorption bands of most chemical and biological molecules as well as fingerprint regions. For practical applications including weak-gas detection [1] and medical diagnostics [2, 3], integrated MIR laser source exhibits enormous prospects thanks to the advantages of small footprint, low manufacturing cost, and strong anti-electromagnetic interference. In recent years, the performance of quantum cascade lasers [4, 5] has been remarkably improved, showing good promise in device integration, which exhibits an up-to-watt-level average power [6, 7] and has a wide spectral tunability range [8]. However, for quantum cascade lasers, there are still certain challenges in the generation of ultrashort pulses [9] and ultra-broadband spectral output [10, 11]. MIR on-chip sources with broad emission spectral bandwidth and ultra-short pulse width based on nonlinear frequency conversion have been extensively investigated, among which, MIR emitters based on third-order nonlinearity ($\chi^{(3)}$) have been mainly studied [12-14]. Broadband MIR supercontinuum with wavelengths extending to 13 μm in SiGe waveguides and frequency combs in Si-micro-resonators has been demonstrated [15]. However, in order to obtain decent MIR output characteristics such as low pump threshold, broadband spectra, or high conversion efficiency, dedicated dispersion engineering or resonators with high-quality factors are usually required. In addition, a moderate-to-small effective mode area is required to enhance the third-order nonlinearity, which usually limits the waveguide dimensions to nanometer sizes. Correspondingly, large mode mismatch and small alignment tolerance induce low coupling efficiency of the nano-waveguide devices, which has become an obstacle in practical applications. To circumvent the above-mentioned limitations, micro-waveguide devices facilitated with strong quadratic nonlinearity ($\chi^{(2)}$) [16, 17] are expected to be an alternative means for simple and efficient MIR generation. Recently, we, for the first time, have experimentally demonstrated a highly efficient single-pass MIR optical parametric generation in a ZnGeP$_2$ (ZGP) micro-waveguide empowered by birefringence phase-matching [18], which opens new pathways to the generation of tunable and efficient MIR lasers from an integrated nonlinear photonics system. In this work, we present a detailed simulation of mode transmission characterizations of the birefringence anisotropic optical waveguide, giving an explicit indication for further waveguide performance optimization.

It is worth noting that mode transmission characteristics of anisotropic waveguides are different from those of isotropic waveguides. When the principal axis of the dielectric tensor is inconsistent with the direction of transmission in anisotropic waveguide, the waveguide modes consist of a combination of ordinary and extraordinary waves. Leakage mode can occur when coupled waves are transmitted in the waveguide but one of the waves does not fully reflect inside the waveguide resulting in power leakage into the cover and substrate regions, which could induce the increase of waveguide mode transmission loss with several orders of magnitude [19]. Hence, waveguide anisotropy characteristics cannot be ignored, and the mode leakage needs to be calculated carefully to optimize the mode transmission performance of the birefringence waveguide device. A number of methods [20-27], such as multilayer approximation [25], perturbation-expansion method [26], and WKB method [27] have been demonstrated to analyze the mode leakage loss in anisotropic waveguides. However, these analysis methods are generally complicated and only valid for weakly guiding anisotropic waveguide platforms. To bridge these obstacles, the finite element method (FEM) and finite difference method (FDM) [28-30] with appropriate perfectly matched layers (PMLs) [31-33] have been proposed for solving full-vector mode transmission of various anisotropic optical waveguide platforms with the advantage of simple formulation and numerical implementation.

In this paper, based on FEM, we, for the first time, systematically and theoretically explore mode transmission characteristics of the presented micrometer ZGP waveguide platform (ZGP-on-SiO$_2$) with considerations of waveguide anisotropy and substrate material absorption. Owning to the high-index contrast between ZGP and the substrate ($\Delta n \sim 2.14$ for ZGP versus air or $\Delta n \sim 1.71$ for ZGP versus SiO$_2$), TM and TE mode transmission loss at a wavelength range of 2 - 12 μm is found to be less than 4 dB/cm and 1.5 dB/cm, respectively, when the phase-matching angle $\theta$ is 48.3º, denoting a remarkable mode leakage loss. Moreover, mode propagation loss variation versus phase-matching angles $\theta$ at interested wavelengths are investigated, showing a non-obvious oscillation of transmission loss, which is different from that in the reported sub-wavelength dimensional weakly guiding waveguide platforms. TM mode crossing phenomenon at some wavelengths occurs which is observed to be related to waveguide anisotropy and could be avoided by choosing an appropriate phase-matching angle. In addition, it is worth noting that TM/TE mode transmission loss variation versus wavelength at a fixed phase-matching angle is consistent with that of substrate absorption variation trends. Hence, ZGP-on-Sapphire waveguide platform performances are also evaluated, pointing toward an improved waveguide design direction.

## 2. Simulation results

To test the feasibility of the employed theoretical FEM model for quantifying mode transmission of the proposed ZGP micro-waveguide, an advanced analysis on a previous example that has been demonstrated in [19] is first performed. The repeated simulation regarding the variation of the modal effective index versus phase-matching angle $\theta$ is shown in Fig. 1, which is nearly identical to that in [19], revealing the validity of the established FEM modal. The ZGP waveguide platform shown in Fig. 2(a) is then characterized in detail with the proven FEM model, while some preliminary results have been reported in [18]. Fig. 2(b) is the scanning electron microscope images of the fabricated waveguide, and the waveguide surface roughness is measured to be less than 0.6μm indicating a remarkable waveguide scattering loss [18].

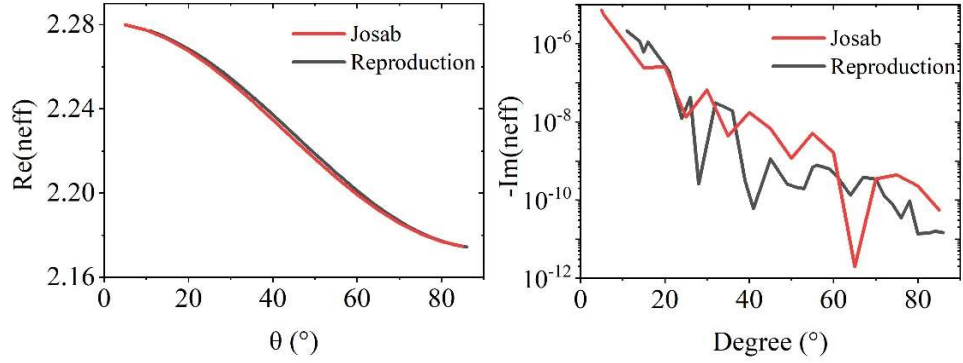

Fig. 1. Real (a) and imaginary part (b) of the modal effective index $n_{eff}$ for the TE mode versus phase-matching angle $\theta$ for the second example in [19]. Our repeated calculation result is marked in red.

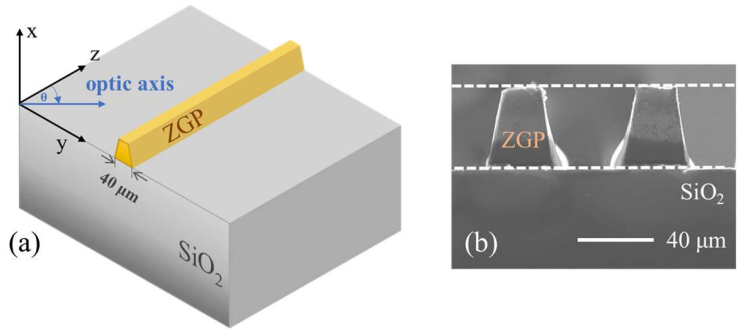

Fig. 2. (a) Schematic diagram of ZGP ridge waveguide. The angle between the optical axis and the propagation direction is denoted as $\theta$. (b) The scanning electron microscope images of the fabricated waveguide.

As mentioned in our previous work [18], a ZGP waveguide designed as 35 and 40 μm in width and height, respectively, with the phase matching angle $\theta$ = 48.3º as depicted in Fig. 2 is employed to perform the broadband and efficient optical parametric generation with an idler output spectrum centered at 8.1 μm, driven at a central wavelength of 2.4 μm. In a further step, to investigate the mode anisotropy, we firstly focus on the variation of modal effective index versus $\theta$ at interested wavelengths including the pump, signal and idler wavelengths at 2.4, 3.4, and 8.1 μm, respectively. Fig. 3(a) shows the real part value variation trends of the modal effective index of TE and TM fundamental modes, corresponding to extraordinary and ordinary waves, respectively, in the ZGP waveguide, which is consistent with the results demonstrated in [34]. Fig. 3(b) shows the corresponding mode transmission loss (dB/cm) taking anisotropy and substrate material absorption into account, calculated from the imaginary part of the modal effective index. The loss profiles are similar to those of the strongly guiding waveguide in Fig. 14(b) in [19], implying that the mode propagation losses are almost independent of phase-matching angles. It is worth noting that only a non-obvious oscillation of mode propagation losses with respect to the phase-matching angle is observed, which is entirely different from those in waveguide structures with weakly guiding [19, 26]. Moreover, the loss values of the pump, signal, and idler waves are confined to $10^{-7}$ dB/cm, $10^{-5}$ dB/cm, and $10^{-1}$ dB/cm respectively, which indicates insignificant mode leaky loss and shows outstanding waveguide performances in the interested wavelengths. In practice, the transmission loss at 2.4 μm pump wavelength in TM polarization of the $\chi^{(2)}$ ZGP waveguide shown in Fig.2(b) is obtained as 0.8 dB/cm with cut-back measurement technique. Efficient MIR light source based on the ZGP waveguide platform is also realized, verifying the effectiveness of demonstrated simulation results [18].

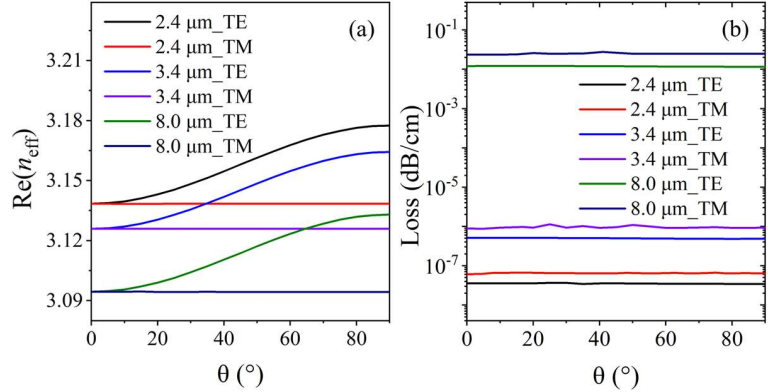

Fig. 3. (a) Real part of the modal effective index of TE and TM fundamental modes versus phase-matching angle $\theta$ at interested wavelengths. (b) Transmission loss of TE and TM fundamental modes versus phase-matching angle $\theta$ at interested wavelengths, which is calculated from the imaginary part of the modal effective index.

Subsequently, the waveguide transmission loss with respect to the laser wavelength in the ZGP-on-SiO$_2$ waveguide platform is investigated, when $\theta$ is set as 48.3º, as shown in Fig. 4(a). It is found that propagation losses of fundamental TE and TM modes at a wavelength shorter than 9 μm are relatively low, which guarantees good transmission of pump and parametric waves in the $\chi^{(2)}$ waveguide. The rise of waveguide losses in the wavelength range of 9 - 11 μm is attributed to the absorption of fused silica substrate peaked at 9.5 μm, as presented in the inset Fig. 4(i) which depicts the absorption coefficient of silica as a function of wavelength [35]. As a comparison, inset Fig. 4(ii) gives the calculated waveguide transmission loss versus wavelength when the waveguide anisotropy factor is not considered. The variation trend of mode propagation loss is similar to that in Fig. 4(a) in which anisotropic character is considered, however, an order of magnitude reduction of the loss value is observed, particularly in a wavelength range of 9-11 μm. This indicates that the waveguide anisotropy-induced mode leaky loss is an essential factor in the procedure of birefringence waveguide design. In addition, mode crossing phenomenon related to anisotropy needs to be considered as shown in Fig. 4(b), which depicts the simulated TM mode profile at representative wavelengths (denoted by dot lines in Fig. 4(a)) 9.6 μm, 9.7 μm, and 9.8 μm, respectively. Notably, mode crossing may occur among waveguide fundamental TM modes and other high-order modes, which could induce a corresponding increase in mode transmission loss [36].

To further explore the TM mode crossing presented in Fig. 4, more simulations on waveguide mode transmission at $\theta$ = 52º and 55º are executed in Fig. 5. Fig. 5(a) shows TM mode profile distribution comparisons of three phase-matching angles at wavelengths of 9.6 μm, 9.7 μm, and 9.8 μm, respectively. The mode crossing phenomenon of these wavelengths at $\theta$ = 48.3º is observed to disappear (i.e. mode de-crossing) by changing the phase-matching angle $\theta$ to 52º and 55º in the designed waveguide, which confirms the necessity and significance of anisotropy characterization in designing the birefringence waveguide. Correspondingly, thanks to the mode de-crossing, TM mode transmission loss of the wavelengths yields an obvious decrease, when the phase-matching angle deviates from 48.3º, as a proof of illustration in Fig. 5(b), which exhibits a transmission loss comparison in the wavelength range of 8-11 μm at different $\theta$. It is worth noting that the mode crossing phenomenon also appears at other wavelengths (dotted blue line in Fig. 5(b)), which is closely correlated to the anisotropy in the birefringence waveguide.

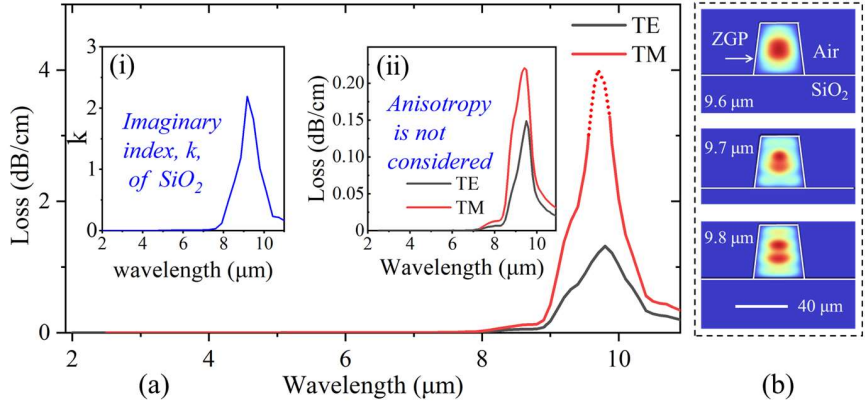

Fig. 4. (a) Transmission losses of fundamental modes in the designed ZGP waveguide in TE and TM polarizations with a phase-matching angle $\theta$ = 48.3º in a broad spectral range of 2 - 11 μm. Insert: (i) absorption coefficient (imaginary part of complex refractive index) of silica, as a function of wavelength. (ii) the calculated waveguide transmission loss in the same condition as Fig. (a), except that the anisotropy characteristic is not considered. (b) The simulated electric field distributions of 9.6 μm, 9.7 μm, and 9.8 μm at $\theta$ = 48.3º in TM mode, showing mode crossing phenomenon.

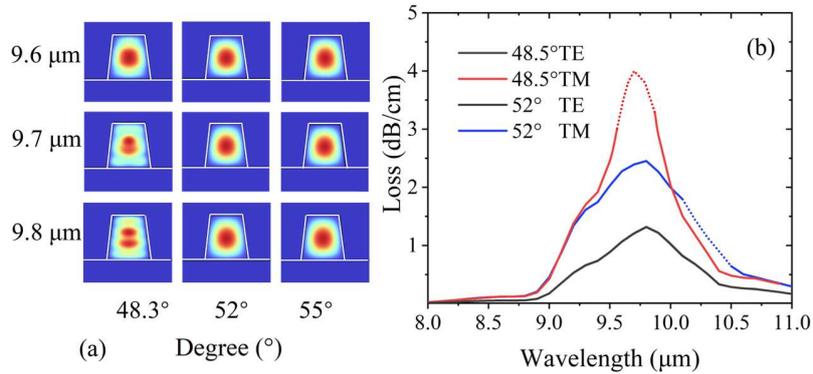

Fig. 5. (a) The simulated electric field diagrams, in TM mode, of 9.6 μm, 9.7 μm, and 9.8 μm at phase-matching angle $\theta$ of 48.3°, 52°, and 55°, respectively. (b) The comparison of calculated transmission loss in TE and TM polarizations between phase-matching angles 48.3º and 52º in the wavelength range of 8 - 11 μm.

SiO$_2$ substrate is employed in our waveguide platforms, which is believed to be an additional factor affecting mode transmission loss. As shown in Fig. 4, a similar variation trend of the waveguide mode transmission loss with that of the material absorption of SiO$_2$ is observed. Therefore, as a proof of demonstration, we also give a mode transmission loss simulation at $\theta$ = 48.3º in another waveguide platform with the employment of sapphire substrate. Fig. 6(a) shows the waveguide loss curve, and the insert figure is the material absorption curve of sapphire. Compared to the ZGP-on-SiO$_2$ platform, the ZGP-on-Sapphire waveguide platform presents much smaller mode transmission loss particularly in the wavelength range of 8 - 11 μm, attributed to a small imaginary index of sapphire, which implies that adopting substrates with low absorption in the long-wavelength infrared region could further improve the performance of the nonlinear waveguide. In addition, we also notice that even using a substrate material of sapphire, mode crossing still occurs at certain wavelengths (dotted red line in Fig. 6(a)), revealing intrinsic anisotropy characteristics of the waveguide structure. Fig. 6(b) also exhibits mode transmission loss as a function of $\theta$ at three wavelengths, which varies within an order of magnitude and is consistent with the conclusion drawn in Fig. 3.

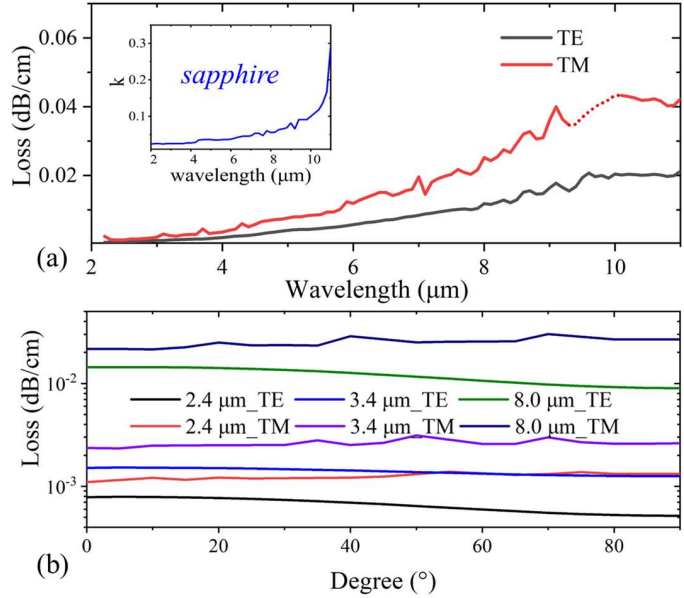

Fig. 6. (a) The calculated propagation loss of TE and TM fundamental modes in a designed waveguide by choosing sapphire as the substrate, with the phase-matching angle $\theta = 48.3°$ in a wavelength range of 2 - 11 μm. Insert: absorption coefficient (imaginary part of complex refractive index) of sapphire, as a function of wavelength. (b) Transmission loss of TE and TM fundamental modes versus the phase-matching angle $\theta$ at the wavelength of 2.4 μm, 3.8 μm, and 8.0 μm.

## 3. Conclusion

In summary, based on the FEM modal, we theoretically investigate mode transmission characteristics of the proposed micro-ZGP waveguide. Waveguide anisotropy is an important factor for designing the waveguide mode transmission and is highly related to the mode crossing phenomenon at certain wavelengths in a birefringence waveguide. It is found that the strong guiding owing to the high-index contrast between core and substrate leads to a remarkable mode leakage. TM and TE mode transmission losses in the wavelength range of 2 - 12 μm are calculated to be less than 4 dB/cm and 1.5 dB/cm, respectively, at the phase matching angle $\theta = 48.3°$. Meanwhile, mode transmission loss variation presents independence with phase matching angle, and non-obvious loss value oscillation is also observed, which ensures design flexibility. Waveguide substrate absorption is an additional decisive factor for waveguide transmission loss. A corresponding performance comparison between ZGP-on-$SiO_2$ and ZGP-on-Sapphire platforms is conducted. The simulation results suggest that waveguide performances could be improved by choosing sapphire as the substrate. Our demonstration also gives an unequivocal indication for other birefringence waveguide designs such as $LiGaS_2$, and $AgS_2$, motivating next-generation MIR-integrated photonics.

**Funding.** National Natural Science Foundation of China (62075144, 62005186, U22A2090), Sichuan Outstanding Youth Science and Technology Talents (2022JDJQ0031), and Engineering Featured team Fund of Sichuan University (2020SCUNG105).